\documentclass[preprint2]{aastex}
\usepackage{epsfig}
\usepackage{natbib}                % To format bibliographies.
\usepackage{lscape}
\usepackage{verbatim}              % Allows quoting source with commands.
\usepackage{graphicx}              % For powerful manipulation of figures.
\usepackage{amsmath,amsthm,amsfonts,amsopn,amssymb} % Some nice math packages.
\usepackage{afterpage}
\usepackage{rotating}					% Allows you to rotate figures, text and tables 
\shorttitle{Astrophysical Parameters and HZ of GJ 581}

\shortauthors{von Braun \& Boyajian et al.}
\begin{document}

%%\received{}
%%\accepted{}

\title{Astrophysical Parameters and Habitable Zone of the Exoplanet Hosting Star GJ 581} 

\author{Kaspar von Braun\altaffilmark{1,11},
Tabetha S. Boyajian\altaffilmark{2,3}, 
Stephen R. Kane\altaffilmark{1},
Gerard T. van Belle\altaffilmark{4},  
David R. Ciardi\altaffilmark{1},  
Mercedes L\'{o}pez-Morales\altaffilmark{5, 9},
Harold A. McAlister\altaffilmark{2}, 
Todd J. Henry\altaffilmark{2}, 
Wei-Chun Jao\altaffilmark{2},
Adric R. Riedel\altaffilmark{2},
John P. Subasavage\altaffilmark{8},
Gail Schaefer\altaffilmark{6},
Theo A. ten Brummelaar\altaffilmark{6}, 
Stephen Ridgway\altaffilmark{7}, 
Laszlo Sturmann\altaffilmark{6}, 
Judit Sturmann\altaffilmark{6}, 
Jude Mazingue\altaffilmark{6},
Nils H. Turner\altaffilmark{6},
Chris Farrington\altaffilmark{6}, 
P. J. Goldfinger\altaffilmark{6}, 
Andrew F. Boden\altaffilmark{10}
}

\altaffiltext{1}{NASA Exoplanet Science Institute, California Institute of Technology, MC 100-22, Pasadena, CA 91125}
\altaffiltext{2}{Center for High Angular Resolution Astronomy and Department of Physics and Astronomy, Georgia State University, P. O. Box 4106, Atlanta, GA 30302-4106} 
\altaffiltext{3}{Hubble Fellow} 
\altaffiltext{4}{European Southern Observatory, Karl-Schwarzschild-Str. 2, 85748 Garching, Germany}
\altaffiltext{5}{Institut de Ci\`{e}ncies de L'Espai (CSIC-IEEC), Campus UAB, Facultat Ci\`{e}ncies, Torre C5 parell 2, 08193 Bellaterra, Barcelona, Spain}
\altaffiltext{6}{The CHARA Array, Mount Wilson Observatory, Mount Wilson, CA 91023}
\altaffiltext{7}{National Optical Astronomy Observatory, P.O. Box 26732, Tucson, AZ 85726-6732}
\altaffiltext{8}{Cerro Tololo Interamerican Observatory, Casilla 603, La Serena, Chile}
\altaffiltext{9}{Department of Terrestrial Magnetism, Carnegie Institution of Washington, 5241 Broad Branch Road, NW, Washington, DC 20015}
\altaffiltext{10}{California Institute of Technology, Pasadena, CA 91125}
\altaffiltext{11}{kaspar@caltech.edu}

\slugcomment{accepted for publication in ApJL}
%%\slugcomment{}
%%\paperid{}

%%%%%%%%%%%%%%%%%%%%%%%%%%%%%%%%%%%%%%%%%%%%%%%%%%%%%%%%%%%%%%

\begin{abstract}

GJ~581 is an M dwarf host of a multiplanet system. We use long-baseline interferometric measurements from the CHARA Array, coupled with trigonometric parallax information, to directly determine its physical radius to be $0.299 \pm 0.010 R_{\odot}$. Literature photometry data are used to perform spectral energy distribution fitting in order to determine GJ~581's effective surface temperature $T_{\rm EFF}=3498 \pm 56$ K and its luminosity $L=0.01205 \pm 0.00024 L_{\odot}$. From these measurements, we recompute the location and extent of the system's habitable zone and conclude that two of the planets orbiting GJ~581, planets d and g, spend all or part of their orbit within or just on the edge of the habitable zone.

\end{abstract}


%%%%%%%%%%%%%%%%%%%%%%%%%%%%%%%%%%%%%%%%%%%%%%%%%%%%%%%%%%%%%%%

\section{Introduction}\label{sec:introduction}

The formation, evolution, and environment of extrasolar planets are heavily dependent on the astrophysical properties of their respective parent stars. In particular, the location and extent of the system's habitable zone (HZ) are direct functions of the parent star's size and surface temperature, both of which are frequently determined by stellar modelling. The advent of long-baseline interferometry at wavelengths in the near-infrared or optical range has made it possible to directly measure these stellar astrophysical properties \citep[e.g., ][and references therein]{bai07,bai08,bai09,bai10,van09}.

M dwarfs are popular targets for exoplanet searches, in part due to their low intrinsic luminosity, which makes it possible for planets with orbital periods of much less than a year to be located near or inside the system's habitable zone. 
%For the transit and radial velocity (RV) planet detection techniques, planets with lower orbital periods are easier to detect due to observational window function considerations \citep{von09}. Furthermore, low stellar mass increases the RV signal amplitude for a given planetary mass, and low stellar radius amplifies transit depth for given planet radius. 
GJ~581 (= HIP~74995, $V=10.570, H=6.095, K=5.837$) is a nearby M2--3 dwarf \citep{bes90a,hen94,haw96,cut03} that hosts six currently reported planets, all detected by the radial velocity technique and successively announced in \citet[][Neptune-mass planet b]{bfd05}, \citet[][super-Earth mass planets c \& d]{udr07}, \citet[][Earth-mass planet e]{mbf09}, and most recently, \citet[][super-Earth mass planets f \& g]{vog10}.
%; see also \citet{ang10}. 
%Planets c and d lie on the inner and outer edge of the system's HZ, and planet g inside it. 

Here we present the fundamental astrophysical parameters of GJ~581 primarily based on interferometric observations, and comment on the resulting location and extent of the system's HZ as well as the locations of the orbiting planets with respect to it. Our observations are described in \S \ref{sec:observations}, and the calculations of the resulting stellar astrophysical parameters in \S \ref{sec:properties}. We discuss the HZ of the GJ~581 system in \S \ref{sec:habitability}, and summarize and conclude in \S \ref{sec:conclusion}.

%%%%%%%%%%%%%%%%%%%%%%%%%%%%%%%%%%%%%%%%%%%%%%%%%%%%%%%%%%%%%%%

\section{Observations}\label{sec:observations}

%%%%%%%%%%%%%%%%%%%%%%%%%%%%%%%%%%%%%%%%%%%%%%%%%%%%%%%%%%%%%%%

\subsection{Interferometry}\label{sec:interferometry}

Our interferometric observations of GJ~581 were carried out using the Georgia State University Center for High Angular Resolution Astronomy (CHARA) Array \citep{ten05} in the context of ongoing surveys to determine fundamental parameters of exoplanet hosting stars \citep{bbc10} and of low-mass stars \citep{boy10}.
%CHARA is a six element optical/IR interferometer located on the grounds of Mount Wilson Observatory \citep{ten05}.  
We used the single baseline CHARA Classic beam combiner in $H$-band ($\lambda_{central} = 1.67$~$\mu$m) and collected data on 6 nights between March and June 2010. 
%(see Table~\ref{tab:observations}) 
We observed with CHARA's longest baselines S1/E1 ($\sim 330$m, 19 observations) and W1/E1 ($\sim 313$m, 8 observations).  The calibrator stars HD~136257 ($V=7.55, H=6.24, \theta_{\rm SED}=0.233\pm0.007$ milli-arcseconds\footnote{$\theta_{\rm SED}$ corresponds to the estimated angular diameter based on spectral energy distribution fitting.}) and HD~136713 ($V=7.99, H=5.83, \theta_{\rm SED}=0.323\pm0.023$ milli-arcseconds) were observed along with GJ~581 in bracketed sequences.  These calibrator stars were chosen to be near-point-like sources of similar $H$-band magnitudes as GJ~581 and located at small angular distances from it. 
%(Table~\ref{tab:calibrators}). 
%Fig. \ref{fig:diameters} shows our interferometric results.

\begin{figure}										%			Visibility plot
\centering
\epsfig{file=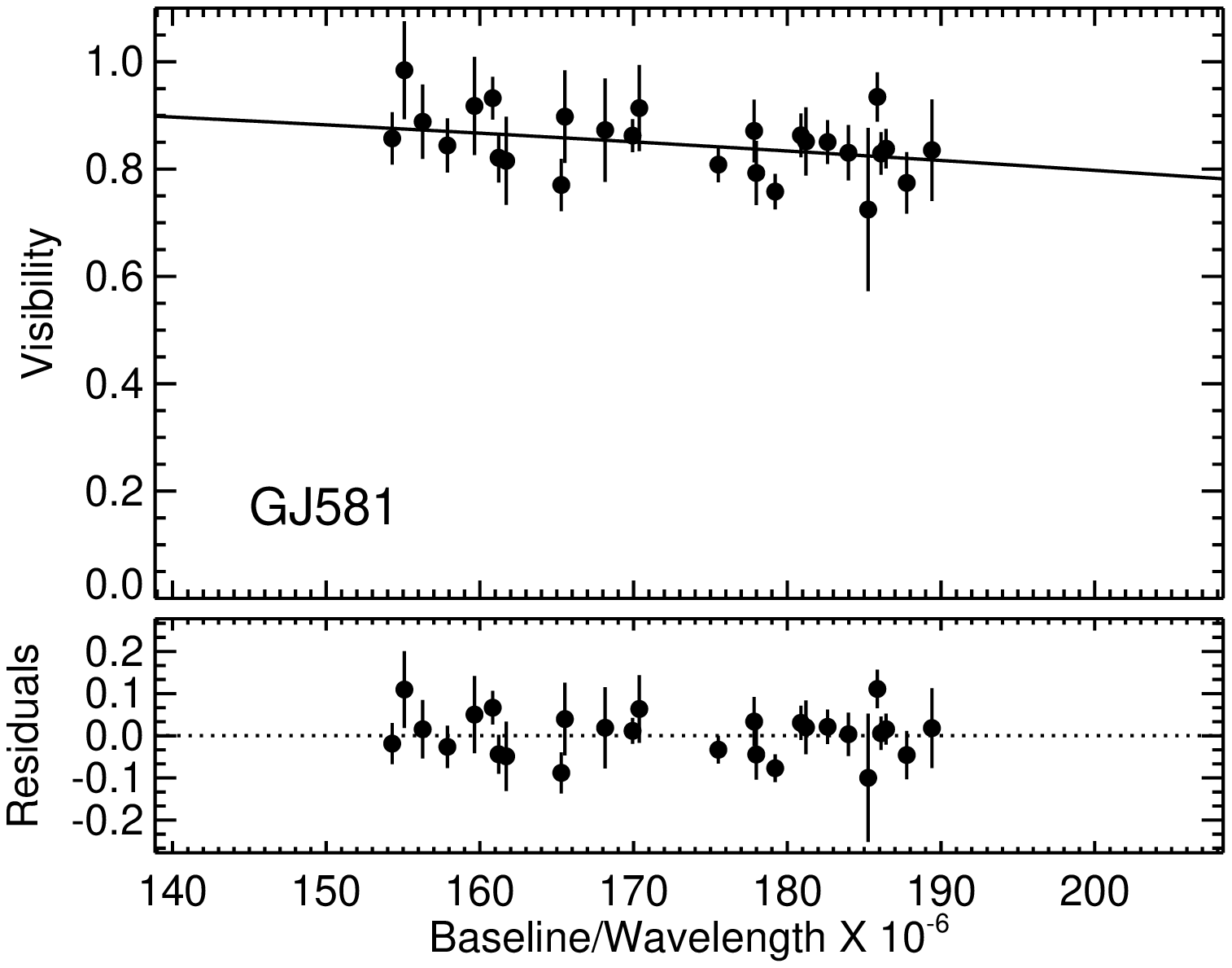,width=0.9\linewidth,clip=} \\
\caption[Angular Diameters] {Calibrated visibility observations along with the limb-darkened angular diameter fit for GJ~581. For details, see \S \ref{sec:interferometry}.}
\label{fig:diameters}
\vspace{.5 cm}
\end{figure}

The uniform disk and limb-darkened angular diameters ($\theta_{\rm UD}$ and $\theta_{\rm LD}$, respectively; see Table \ref{tab:properties}) are found by fitting the calibrated visibility measurements (Fig. \ref{fig:diameters}) to the respective functions for each relation.  These formulae may be described as $n^{th}$-order Bessel functions that are dependent on the angular diameter of the star, the projected distance between the two telescopes and the wavelength of observation (see equations 2 and 4 of \citealt{han74}).  We use the $H$-band limb-darkening coefficient ($\mu_H = 0.358$) calculated from the PHOENIX model atmospheres code for the corresponding value of $T_{\rm EFF} = 3500$~K and $\log g =5.0$, as tabulated in \citet{cla00}\footnote{We note that an uncertainty in an assumed parameter will only modestly influence the value of the coefficient, and consequentially the value of $\theta_{\rm LD}$.  For example, changing the model $T_{\rm EFF}$ by $\pm 300$~K will change the resulting diameter by only a few tenths of a percent, well within our error budget.}. The solutions and their respective errors are computed using MPFIT, a non-linear least-squares fitting routine in IDL \citep{mar09}.  For this solution we find $\chi^{2}_{reduced}$ = 0.65, implying that the standard CHARA reduction pipeline may have overestimated our measurement errors. Thus the results presented here assume a true $\chi^{2}_{reduced}$ = 1 to remedy effects of the overestimated uncertainties. Our results are shown in Table \ref{tab:properties}.

Finally, our interferometric data allowed for a search for separated fringe packets to comment on the potential existence of a previously unresolved stellar companion in a near face-on orbit (see discussion in \citealt{bai10,far10}), which would influence the interpretation of the RV curves.  The fringe packets from two stellar sources would overlap for small projected separations of $\sim$0.5--5 milli-arcseconds (mas), producing a periodic modulation in the angular diameter visibility curve (e.g. \citealt{bai10}).  The amplitude of this modulation is set by the flux ratio of the binary components.  Based on the range of visibility residuals in our angular diameter fit, we can rule out companions with magnitude differences $\delta H <$ 2.4 mag. The fact that we did not find any evidence of a secondary fringe packet (and thus stellar-mass companion at low inclination angles) reinforces the dynamical stability argument of coplanar orbits in this system as described in \citet{mbf09} and \citet{bbd08} by providing an empirical lower-limit constraint on the orbital inclinations of the planetary orbits. 

%%%%%%%%%%%%%%%%%%%%%%%%%%%%%%%%%%%%%%%%%%%%%%%%%%%%%%%%%%%%%%%

\subsection{Astrometry}\label{sec:astrometry}

Translating the measured angular radius of GJ 581 into a linear radius depends on its distance.  Trigonometric parallax measurements are available for GJ~581 in {\it The General Catalogue of Trigonometric Stellar Parallaxes} \citep[$\pi_{trig}$ = 157.9 $\pm$ 5.6 mas]{van95}, from van Leeuwen's reduction of the {\it HIPPARCOS} space mission observations \citep[$\pi_{trig}$ = 160.91 $\pm$ 2.62 mas]{vanl08}, and from the RECONS group\footnote{www.recons.org} of $\pi_{trig}$ = 160.01 $\pm$ 1.61 mas\footnote{This supersedes the value reported in \citet{jao05}, where details about the RECONS astrometry program at the CTIO 0.9m can be found.  In essence, the improved parallax results from a reduction of 224 total frames of the GJ~581 field taken over 9.93 years, whereas the previous result was based on 122 frames over 2.95 years.}.  The weighted-mean parallax of the three independent measurements for the GJ~581 system is 160.12 $\pm$ 1.33 mas, driven primarily by the improved RECONS parallax, and is used here in the calculation of GJ~581's physical size.

%%%%%%%%%%%%%%%%%%%%%%%%%%%%%%%%%%%%%%%%%%%%%%%%%%%%%%%%%%

\section{Fundamental Astrophysical Parameters of the Host Star GJ~581}\label{sec:properties}

A summary of our results of the fundamental parameters of GJ~581 is given in Table \ref{tab:properties}. We describe below how individual values were obtained.

\begin{figure}										%			SED plot
\centering
\epsfig{file=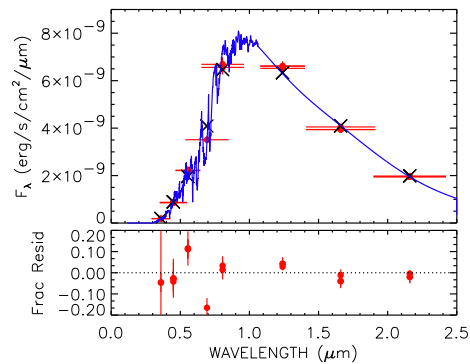,width=0.9\linewidth,clip=} \\
\caption[SED Fit] {Spectral energy distribution fit for GJ~581. Literature photometry data are plotted on top of the spectral template. Horizontal error bars correspond to filter bandwidths. Crosses indicate flux levels of the spectral template integrated over the filter bandwidth. The fractional residuals are given in the bottom panel, along with scaled error bars in the respective photometry datapoint. The large $U$-band error bar is due to uncertainties associated with literature photometry. For details, see \S \ref{sec:properties} and Table \ref{tab:properties}.}
\label{fig:sed}
\vspace{.5 cm}
\end{figure}

Based on GJ~581's $\pi_{trig}$ and interferometrically measured angular diameter $\theta_{\rm LD}$, we calculate its linear radius to be $R = 0.299 \pm 0.010 R_{\rm \odot}$. Following the procedure outlined in \S3.1 of \citet{vcb07}, we produce a fit of the stellar spectral energy distribution (SED) to literature $UBVRIJHK$ photometry. We obtain a value of GJ~581's stellar bolometric flux $F_{\rm BOL} = (9.92 \pm 0.104)\times10^{-9}$~erg cm $^{-2}$ s$^{-1}$  and luminosity $L = 0.01205 \pm 0.00024 L_{\odot}$. Photometry for our SED fit from the following papers was used along with spectral templates from \citet{pick98}: \citet{wei93}, \citet{bes90a}, \citet{koe02}, \citet[][2MASS $JHK_s$]{cut03}, \citet{nor04}, \citet{kil07}, \citet{ray09}, and \citet{koe10}. The transformation from the SAAO photometry system to 2MASS was based on the equations in \citet{koe07}. We show our SED fit in Fig. \ref{fig:sed}.

It should be noted that, in order to minimize the $\chi^2$ of our SED fit, we obtain a non-zero value for extinction ($A_V = 0.174 \pm 0.021$), which is unexpected for a star at a distance of around 6 pc. Forcing $A_V=0$ degrades the $\chi^2$ from around 2 to around 6. The associated difference in the calculated value of $F_{\rm BOL}$ is around 6.7\%, corresponding to a 1.6\% effect in the calculation of $T_{\rm EFF}$ (see Equation \ref{eq:temperature} below) in the sense that an $A_V>0$ produces a slightly higher $T_{\rm EFF}$. We postulate that the extinction could be circumstellar in origin\footnote{Both \citet{kos09} and \citet{bbc09} report a 70 $\mu$m excess in the analysis of {\it Spitzer} MIPS archival data of around three times the value expected from extrapolating the stellar SED to wavelengths redward of 2MASS photometry, although \citet{bbc09} classify the significance of their detection as marginal.}. Equivalent SED fitting using NextGen synthetic templates \citep{hau99} instead of the empirical \citet{pick98} templates yielded near-identical values of $F_{\rm BOL}$ and $A_V$ for the best-fit templates using $T_{\rm EFF}=3300$ K, log g = 5.0, [Fe/H] = -0.5.  
%This value of [Fe/H] is consistent with the sub-solar value of -0.25 found in \citet{but06}.

The effective temperature $T_{\rm EFF}$ is calculated based on the re-arranged form of the Stephan-Boltzmann equation

\begin{equation} \label{eq:temperature}
T_{\rm EFF} ({\rm K}) = 2341 (F_{\rm BOL}/\theta_{\rm LD}^2)^{\frac{1}{4}},
\end{equation}

\noindent
where $F_{\rm BOL}$ is in units of $10^{-8}$~erg cm$^{-2}$ s$^{-1}$ and the angular diameter $\theta_{\rm LD}$ is in units of mas. We calculate GJ~581's effective temperature to be $T_{\rm EFF} = 3498 \pm 56$~K. 

We derive a mass estimate of $M \simeq 0.3 M_{\rm \odot}$ from the $K$-band Mass-Luminosity relation in \citet{del00}. Finally, GJ~581's mass and radius produce $\log g = 4.96 \pm 0.08$, assuming a mass uncertainty of 0.05 $M_{\odot}$. See Table~\ref{tab:properties} for a summary of our results.

%Using GJ~581's $\pi_{trig}$ (\S \ref{sec:astrometry}) and $K$ magnitude (\S \ref{sec:introduction}), we 

The most recent values for GJ~581's astrophysical properties before the ones presented here are the ones calculated in \citet{bfd05}. They were subsequently adopted by the papers documenting the discoveries of the other five planets orbiting GJ~581 \citep{udr07,mbf09,vog10}. Below we compare these values to ours. 

%, and they agree well with these measured quantities presented in this work.  Our value of the absulute luminosity $L = 0.01212\pm 0.00027 L_{\rm \odot}$ was directly determined, incorporating the new and improved parallax measurement and SED determineation of the $F_{\rm BOL}$, and it agrees with the estimated luminosity from 

\citet{bfd05} use a $V$-band bolometric correction from \citet{del98} to derive $L = 0.013 L_{\rm \odot}$, which constitutes a 4$\sigma$ increase with respect to our value, using only our error estimates; theirs are not given. This discrepancy decreases only marginally (to 3.6$\sigma$) if we set equal the two slightly different values for $\pi_{trig}$.  
%This estimate is $\sim$2\% lower than the one presented in \citet{bfd05}, $M = 0.31 M_{\rm \odot}$, derived using the same relation. 
The radius of GJ~581 was calculated in \citet{bfd05} by applying their estimated mass, $M = 0.31 M_{\rm \odot}$, to the Mass-Radius relation in \citet{cha00}. We find that their resulting value of 0.29 $R_{\odot}$ is $\sim$1$\sigma$ (or $\sim$3.0\%) smaller than our direct measurement. Finally, the estimate of $T_{\rm EFF}$ in \citet{bfd05}, based on their calculated stellar radius and luminosity, is 3622~K, which is 3.5\% (2.2$\sigma$) higher than our value.
% (again based only on our error estimates). 

Interestingly, some previous estimates of GJ~581's $T_{\rm EFF}$ yield results with lower temperature estimates. The temperature calibration for M dwarfs presented in \citet{bes95} implies a value for GJ~581's $T_{\rm EFF}$ of 3320 K. Furthermore, two different approaches used in \citet{cas08} result in $T_{\rm EFF}$=3320 K and $T_{\rm EFF}$=3300 K, respectively. Conversely, the (independently calculated) empirical relation  between $T_{\rm EFF}$ and $(V-K)_0$ in \citet[][equation 2]{van09} produces an effective temperature of GJ~581 of just over 3500 K. 
  
% is this star later than a M2-3?? like the V-K of almost 5 would have you believe? older (<1970) spectral classifications list older spectral types, but now its a spectral standard for an early M2.5?  ask Todd about his paper on this.
   
%   * the vogt paper section 3 says delfosse et al 1998 BC gives a L= 0.013 Lsun.  This is ~2.5 sigma brighter than what we get (0.012 +/0.00043), thus leading them to more massive star when applying mass-Luuninosity relation
   
%selsis et al 2007: page 30 of vogt et al 2010

%%%%%%%%%%%%%%%%%%%%%%%%%%%%%%%%%%%%%%%%%%%%%%%%%%%%%%%%%%

\section{The Habitable Zone in the GJ~581 System}\label{sec:habitability}

%The HZ is defined as the range of circumstellar distances from a star within which a planet could have liquid water on its surface, given a dense enough atmosphere \citep{kas93,und03,skl07,jon10}. 

%Stephen: orbital parameters from vogt paper (to keep everything consistent). g planet eccentricity assumed to be zero, hence the intersection of the orbits.

%references: 

%Some relevant papers to cite (plus look into their citations): 
%\citet{mbf09},
%\citet{za09},
%\citet{lmb06},
%\citet{bbc07},
%\citet{bfd05},
%\citet{skl07},
%\citet{bdf07},
%\citet{bbd08}.

%{\it Kaspar with David for at least the age estimates}
%time scale of planet formations and debris disk formations.

%\section{The Habitable Zone in the GJ~581 System}
%\label{sec:habitability}

A system's traditional HZ is defined as the range of circumstellar
distances from a star within which a planet could have liquid water on
its surface, given a dense enough atmosphere. The various criteria for
defining the HZ are described in detail in \citet{kas93} and
further generalized by \citet{und03}. For the GJ~581 system, the HZ
boundaries were calculated by \citet{udr07}, \citet{mbf09}, and
\citet{vog10} in the context of the respective detections of additional planets in the
GJ~581 system. A detailed analysis of the habitablility of the system under
various assumptions can be found in \citet{skl07}. 
%Here we
%recalculate HZ boundaries for the newly measured host star
%properties.

Our recalculation of the boundaries of the GJ~581's HZ based on our directly determined host star properties is based on the equations in \citet{und03} and \citet{jon10} that relate the radii of the inner and outer edges of the HZ to the luminosity and effective temperature of the host star. This results in an inner boundary of 0.11 astronomical units (AU) and an outer boundary of slightly above 0.21~AU. To determine which of GJ~581's planets are located in the HZ, we calculate their equilibrium temperatures $T_{eq}$ following the methods of \citet{skl07}:
\begin{equation}\label{eq:equitemp}
  T_{eq}^4 = \frac{S (1 - A)}{f \sigma},
\end{equation}
where $S$ is the stellar energy flux, $A$ is the Bond albedo, and $\sigma$ is the Stefan-Boltzmann constant. The redistribution factor $f$ is determined by the efficiency of atmospheric heat redistribution efficiency and is set to 2 for a hot dayside and to 4 for even heat redistribution. Table \ref{tab:equiltemp} shows the calculated equilibrium temperatures for the planets in the GJ~581 system assuming an Earth Bond albedo of 0.29 in each case. We look at two different scenarios of the geometry of the GJ~581 system below. 

\begin{figure*}										% 			HZ plots
  \begin{center}
    \begin{tabular}{cc}
      \epsfig{file=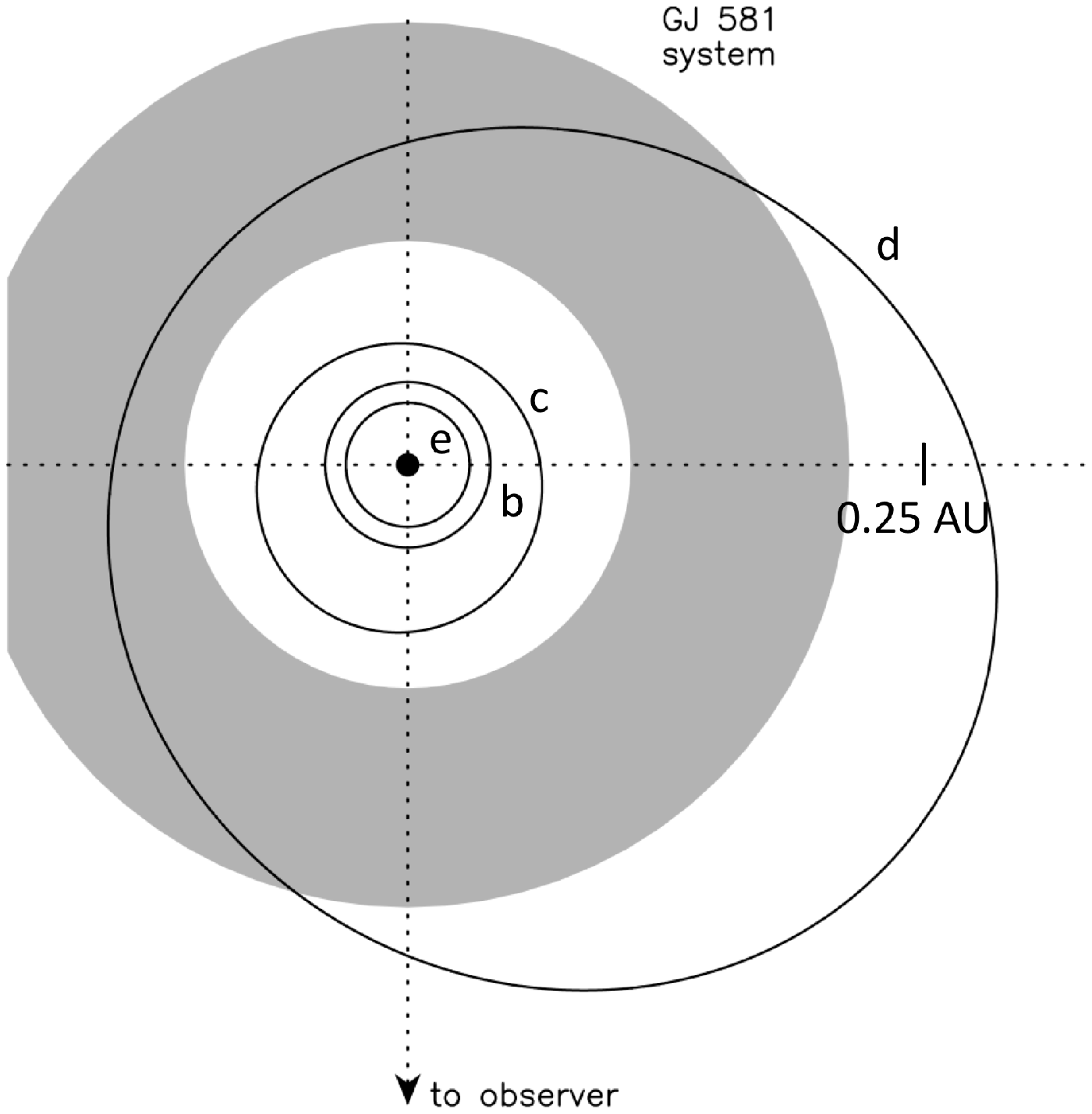,width=0.5\linewidth,clip=} &
      \epsfig{file=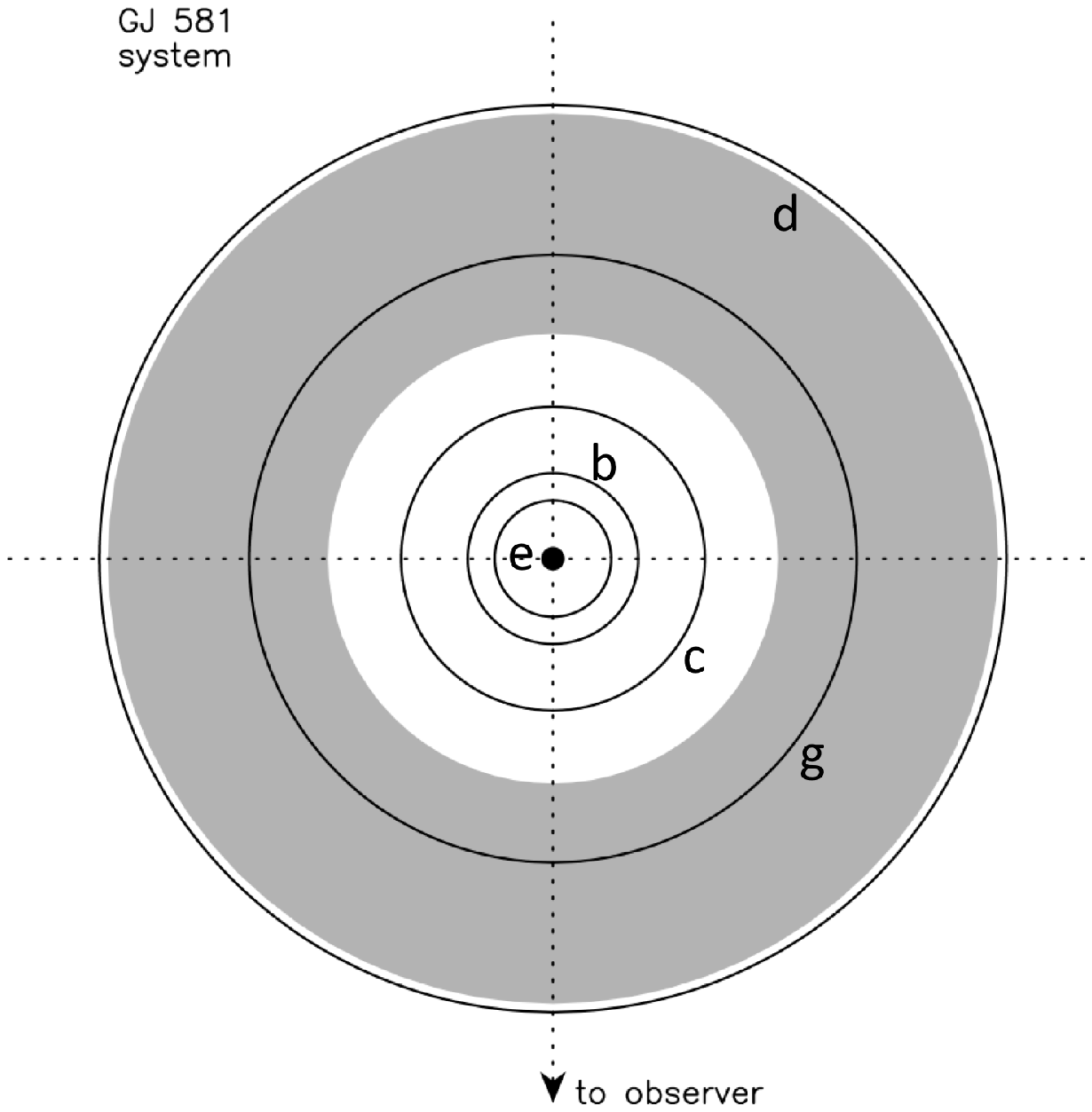,width=0.5\linewidth,clip=} \\
    \end{tabular}
  \end{center}
  \caption{A top-down view of the GJ~581 system. The habitable zone is
      indicated by the gray shaded region with calculated boundaries of 0.11 and 0.21 AU. Left panel: scenario 1 adopted from \citet{mbf09}. Planet d spends part of its elliptical orbit in the HZ. Right panel: scenario 2 adopted from \citet{vog10}. Planet g spends all its orbits inside the HZ, which planet d orbits right on its outer edge. Planet f ($a$ = 0.758 AU) is not shown for purpose of clarity. For details, see \S \ref{sec:habitability} and Table \ref{tab:equiltemp}.}
  \label{fig:hz_orbits}
\end{figure*}

{\bf Scenario 1:} We take orbital parameters from \citet{mbf09}.
% %and \citet[][planets f \& g]{vog10}. Note that \citet{vog10} force zero eccentricity for planets f and g, which results in an intersection of the orbits of planets d and g, as projected onto the plane of the sky. 
Figure \ref{fig:hz_orbits} (left panel) depicts this scenario of the geometry of the GJ~581 system with its HZ shown as a gray shaded region. Planets d 
%and g 
spends part of its orbit in the HZ. 
It should be noted, that due to the non-zero orbital eccentricity, $T_{eq}$ is a function of time (or phase angle). Thus, the $T_{eq}^{f=4}$ for planet d varies from 229~K at periastron to 154~K at apastron, causing it to periodically dip into and out of the HZ.

{\bf Scenario 2:} We take all orbital parameters from \citet{vog10}, motivated by the discussion in \citet{ang10}, which points out that the RV signal of planet d's eccentric orbit could, in fact, be due to two circular orbits of planets d and g. The geometry of this scenario is illustrated in \citet[][figure~6 and table~2]{vog10} and shown here in the right panel of Figure \ref{fig:hz_orbits}. In this scenario, planet g spends all its their circular orbit in the HZ with a semi-major axis of  $a_g = 0.15$~AU, while planet d is right on the outer edge of the HZ with $a_d = 0.22$~AU. 
For planet g, an even heat redistribution results in $T_{eq} = 222$~K which is slightly less than the value calculated by \citet{vog10} as anticipated from our smaller value for GJ~581's luminosity. We note that $T_{eq} = 222$~K is below the freezing temperature of water, but does not take into account the greenhouse effect heating due to an atmosphere (for comparison: Earth's $T_{eq} = 255$~K). See section 6 of \citet{vog10} and particularly \citet{wor10} for more detailed elaborations. Planet d's $T_{eq} = 181$ K for $f=4$ when using its semi-major axis in Equation \ref{eq:equitemp}.

% XXX this is wrong XXX Orbital distances shown in Fig. \ref{fig:hz_orbits} are lower limits and need to be divided by ${\rm sin} i$ to obtain actual distances, where $i$ is the unknown system inclination angle, constrained to $i>40^{\circ}$ in \citet[][see also \S \ref{sec:interferometry}]{mbf09}. This constraint imposes a correction factor between 1--1.56. The discussion below assumes ${\rm sin}i = 1$.

% XXX this is wrong XXX The fraction of its orbit spent within the HZ is dependent on the aforementioned ${\rm sin}i$. For low values ${\rm sin}i$, that fraction would be very small, but we point out that then, an increasing part of planet c's elliptical orbit would be located within the HZ. 

% XXX this is wrong XXX It should also be pointed out that planet g's circular orbit remains fully within the HZ for all values of $1 < {\rm sin}i < 1.56$. 

%%%%%%%%%%%%%%%%%%%%%%%%%%%%%%%%%%%%%%%%%%%%%%%%%%%%%%%%%%%%%%%%%%%%%%%

\section{Summary and Conclusion}    \label{sec:conclusion}

In this paper, we present directly determined astrophysical parameters of the late-type, multiplanet host star GJ~581, observed as part of our ongoing interferometric survey with the CHARA Array of KM dwarfs with and without planets. We find small but significant differences between literature values and our empirical results for GJ~581, which are explicitly shown in \S \ref{sec:properties} and Table \ref{tab:properties}. We measure a stellar radius that is larger than predicted by theory-based calibrations (3\%) but a slightly lower than predicted $T_{\rm EFF}$ (3.5\%), resulting in our calculated luminosity's being lower than the currently used literature value (by $\sim$7\%). Thus, we confirm the results in \citet{boy10} that theoretically calculated radii of M dwarfs are smaller than the directly measured counterparts. 

We recalculate the boundaries of the system's HZ based on our new results. An analysis of the equilibrium temperatures of the six planets orbiting GJ~581 confirms that planet g spends all of its orbit inside the HZ. Depending on the assumed eccentricity of planet d's orbit, it periodically dips into and out of the HZ or spends all of its circular orbit on its outer edge (\S \ref{sec:habitability}). The presence of any moderately dense atmosphere around planets d and particularly g could provide sufficient greenhouse heating of the planetary surface temperatures \citep{skl07,wor10} such that any existing water could be in liquid form.

Planetary characterization is playing an increasingly important role in exoplanet research. With the exception of directly imaged planets, the large majority of astrophysical exoplanet parameters are functions of stellar parameters, which in turn are, by necessity, calculated on the basis of stellar models. Particularly in the low-mass regime, however, these stellar models lack observational constraints, leading, not necessarily by fault of the models themselves, to uncertainties and sometimes systematic discrepancies between theoretical and observational results, as we present in this paper. It is thus hard to overstate the importance of ``understanding the parent stars''. With recent and ongoing improvements in both sensitivity and spatial resolution of near-infrared and optical interferometric data quality, we are able to ameliorate this situation and provide firm, direct measurements of stellar radii and effective temperatures in the low-mass regime. 
%Additional results coming soon!

%%%%%%%%%%%%%%%%%%%%%%%%%%%%%%%%%%%%%%%%%%%%%%%%%%%%%%%%%%%%%%%

\acknowledgments

We would like to express our gratitude to the (semi-)anonymous referee whose comments on GJ~581's surface temperature and on limb-darkening coefficients clearly improved the quality of this manuscript. TSB acknowledges support provided by NASA through Hubble Fellowship grant \#HST-HF-51252.01 awarded by the Space Telescope Science Institute, which is operated by the Association of Universities for Research in Astronomy, Inc., for NASA, under contract NAS 5-26555.  The CHARA Array is funded by the National Science Foundation through NSF grants AST-0606958 and AST-0908253 and by Georgia State University through the College of Arts and Sciences, as well as the W. M. Keck Foundation. This research made use of the SIMBAD literature database, operated at CDS, Strasbourg, France, and of NASA's Astrophysics Data System. This publication makes use of data products from the Two Micron All Sky Survey, which is a joint project of the University of Massachusetts and the Infrared Processing and Analysis Center/California Institute of Technology, funded by the National Aeronautics and Space Administration and the National Science Foundation. This research made use of the NASA/IPAC/NExScI Star and Exoplanet Database, which is operated by the Jet Propulsion Laboratory, California Institute of Technology, under contract with the National Aeronautics and Space Administration.

%%%%%%%%%%%%%%%%%%%%%%%%%%%%%%%%%%%%%%%%%%%%%%%%%%%%%%%%%%%%%%%

% Bibliography %

%\clearpage

%\bibliographystyle{apj}            % Please learn to use the
                                    % formatting of Latex's Bibtex. It
                                    % will make your life easier.
% apj.bst should be in this directory as well as apj-jour.bib and reference paper.bib

%\bibliography{apj-jour,paper}      % "paper.bib" contains all my
                                    % references. "apj-jour.bib"
                                    % contains abbreviations of
                                    % journals.

%%%%%%%%%%%%%%%%%%%%%%%%%%%%%%%%%%%%%%%%%%%%%%%%%%%%%%%%%%%%%%

%%%%%%%%%%%%%%%%%%%%%%%%%%%%%%%%%%%%%%%%%%%%%%%%%%%%%%%%%%%%%%%
%			FIGURES
%%%%%%%%%%%%%%%%%%%%%%%%%%%%%%%%%%%%%%%%%%%%%%%%%%%%%%%%%%%%%%%

%%%%%%%%%%%%%%%%%%%%%%%%%%%%%%%%%%%%%%%%%%%%%%%%%%%%%%%%%%%%%%%

%%%%%%%%%%%%%%%%%%%%%%%%%%%%%%%%%%%%%%%%%%%%%%%%%%%%%%%%%%%%%%%

%%%%%%%%%%%%%%%%%%%%%%%%%%%%%%%%%%%%%%%%%%%%%%%%%%%%%%%%%%%%%%%
%			TABLES
%%%%%%%%%%%%%%%%%%%%%%%%%%%%%%%%%%%%%%%%%%%%%%%%%%%%%%%%%%%%

%\input{observations.tex} % observational info

%%%%%%%%%%%%%%%%%%%%%%%%%%%%%%%%%%%%%%%%%%%%%%%%%%%%%%%%%%%%%%%

%\input{calibrators.tex} % calibrator info

%%%%%%%%%%%%%%%%%%%%%%%%%%%%%%%%%%%%%%%%%%%%%%%%%%%%%%%%%%%%%%%

%\input{gj581_properties.tex} % info on GJ581 host star

\begin{deluxetable}{rcc}

\tablecaption{Stellar Properties of GJ 581 \label{tab:properties}} 
\tablewidth{0pc}
\tablehead{
\colhead{Parameter} &
\colhead{Value} &
\colhead{Reference}	
}
\startdata
Spectral Type \dotfill	&	M2.5V	&	\citet{hen94} 	\\ % hawley 1996 gives M3V spectral type
$V-K$ \dotfill			&	$4.733\pm0.083$	&	\citet{bes90a, cut03}	\\
$[$Fe/H$]$ \dotfill		&	\phn\phn$-0.135$\phs	&	Average\tablenotemark{a}	\\
$\theta_{\rm UD}$ (mas) \dotfill		&	$0.433 \pm 0.014$	&	this work	\\
$\theta_{\rm LD}$ (mas)	\dotfill	&	$0.446 \pm 0.014$	&	this work	\\
Parallax (mas) \dotfill				&	$160.12 \pm 1.33$\phn\phn	&	this work	\\
Radius ($R_{\rm \odot}$) \dotfill	&	$0.299 \pm 0.010$	&	this work	\\
Luminosity ($L_{\rm \odot}$) \dotfill	& $0.01205 \pm 0.00024$	&	this work	\\
$T_{\rm EFF}$ (K)	\dotfill			&	$3498 \pm 56$\phn\phn		&	this work	\\
%Mass ($M_{\rm \odot}$)\tablenotemark{b} \dotfill	&	\phn\phn$0.305$	&	this work \& \citet{del00}	\\
$\log g$\tablenotemark{b}	\dotfill			&	$4.96 \pm 0.08$		&	this work	\\	%Mass ($M_{\rm \odot}$)$^{}$ \dotfill	&	\nodata	&	this work	\\ % no log g in literature
\enddata
\tablecomments{Properties of GJ~581. For details, see \S \ref{sec:properties}.}
\tablenotetext{a}{Average from \citet{bfd05, joh09, roj10, sch10}.}
%\tablenotetext{b}{Mass derived from $K$-band Mass-Luminosity (M-L) relation; no uncertainty quoted for fit.}
\tablenotetext{b}{The surface gravity $\log g$ is based on our quoted values of mass and radius, and its uncertainty is calculated from our empirically determined radius uncertainty. We assume $\sigma M = 0.05 M_{\rm \odot} \simeq$ 16\%, which represents a slightly more conservative approach than the $(\sigma M)_{max}$ = 10\% requirement in the calculation of the M-L relation in \citet{del00}. }
\end{deluxetable}

%%%%%%%%%%%%%%%%%%%%%%%%%%%%%%%%%%%%%%%%%%%%%%%%%%%%%%%%%%%%%%%

%\begin{table}
%  \begin{center}
%    \caption{Equilibrium temperatures for GJ~581 system planets.}
%    \label{tab:equiltemp}
%    \begin{tabular}{@{}cccc}
%      \hline
%      Planet & $a$ (AU) & $T_{eq}^{f=4}$ (K) & $T_{eq}^{f=2}$ (K) \\
%      \hline
%      b & 0.041 & 418 & 498 \\
%      c & 0.070 & 320 & 381 \\
%      d & 0.220 & 181 & 215 \\
%      e & 0.030 & 489 & 582 \\
%      f & 0.758 &  97 & 116 \\
%      g & 0.146 & 222 & 264 \\
%      \hline
%    \end{tabular}
%  \end{center}
%\end{table}

\begin{deluxetable}{cccc}
    \tablecaption{Equilibrium temperatures for the GJ~581 system planets.\label{tab:equiltemp}}    
	\tablewidth{0pc}
    \tablehead{
      \colhead{Planet} & \colhead{$a$ (AU)} & \colhead{$T_{eq}^{f=4}$ (K)} & \colhead{$T_{eq}^{f=2}$ (K)}}
      \startdata
     b & 0.041 & $418 \pm 3$\phn\phn & $498 \pm 3$\phn\phn \\
     c & 0.070 & $320 \pm 2$\phn\phn & $381 \pm 2$\phn\phn \\
     d & 0.220 & $181 \pm 1$\phn\phn & $215 \pm 2$\phn\phn \\
     e & 0.030 & $489 \pm 2$\phn\phn & $582 \pm 3$\phn\phn \\
     f & 0.758 & $ 97 \pm 1$\phn & $116 \pm 1$\phn\phn \\
     g & 0.146 & $222 \pm 2$\phn\phn & $264 \pm 2$\phn\phn \\
%      b & 0.041 & 418 & 498 \\
%      c & 0.070 & 320 & 381 \\
%      d & 0.220 & 181 & 215 \\
%      e & 0.030 & 489 & 582 \\
%      f & 0.758 &  97 & 116 \\
%      g & 0.146 & 222 & 264 \\
      \enddata

\tablecomments{Equilibrium temperatures based on the equations in \citet{skl07} and assuming orbital element values from \citet[][planets b--e]{mbf09} and \citet[][planets f and g]{vog10}. $f = 4$ implies perfect energy redistribution efficiency on the planetary surface, $f = 2$ means no energy redistribution. For details, see \S \ref{sec:habitability} and Fig. \ref{fig:hz_orbits}. Note that the quoted uncertainties are based only on the uncertainties in the calculation of GJ~581's luminosity, and not on uncertainties in the orbital elements. As such, they should be regarded as lower limits.}
\end{deluxetable}

%%%%%%%%%%%%%%%%%%%%%%%%%%%%%%%%%%%%%%%%%%%%%%%%%%%%%%%%%%%%%%%

\end{document}